\documentclass[aip,apl,amsmath,amssymb,reprint,twocolumn,groupeaddress,superscriptaddress,floatfix]{revtex4-1}

\usepackage[latin1]{inputenc}
\usepackage[T1]{fontenc}

\usepackage{amsmath,amssymb}
\usepackage{graphicx}
\graphicspath{{fig/}}

\usepackage{units}
\usepackage{journames}

\usepackage[final]{neel}

\begin{document}

\renewcommand\topfraction{0.8}
\renewcommand\bottomfraction{0.7}
\renewcommand\floatpagefraction{0.7}

\title{Phase diagram of magnetic domain walls in spin valve nano-stripes}%

\author{N.~Rougemaille}
\affiliation{Institut N\'{E}EL, CNRS \& Universit\'{e}
Joseph Fourier -- BP166 -- F-38042 Grenoble Cedex 9 -- France}%
\author{V.~Uhl\'\i \v{r}}
\affiliation{CEITEC BUT, Brno University of Technology, Technick\'{a} 10, 61669 Brno, Czech Republic}
\affiliation{Institut N\'{E}EL, CNRS \& Universit\'{e}
Joseph Fourier -- BP166 -- F-38042 Grenoble Cedex 9 -- France}%
\author{O.~Fruchart}
\affiliation{Institut N\'{E}EL, CNRS \& Universit\'{e}
Joseph Fourier -- BP166 -- F-38042 Grenoble Cedex 9 -- France}%
\author{S.~Pizzini}
\affiliation{Institut N\'{E}EL, CNRS \& Universit\'{e}
Joseph Fourier -- BP166 -- F-38042 Grenoble Cedex 9 -- France}%
\author{J.~Vogel}
\affiliation{Institut N\'{E}EL, CNRS \& Universit\'{e}
Joseph Fourier -- BP166 -- F-38042 Grenoble Cedex 9 -- France}%
\author{J.~C.~Toussaint}
\affiliation{Institut N\'{E}EL, CNRS \& Universit\'{e}
Joseph Fourier -- BP166 -- F-38042 Grenoble Cedex 9 -- France}%
\affiliation{Grenoble - Institut National Polytechnique -- France}%

\date{\today}%

\begin{abstract}

We investigate numerically the transverse versus vortex phase diagram of head-to-head domain walls in Co/Cu/Py spin valve nano-stripes~(Py: Permalloy), in which the Co layer is mostly single domain while the Py layer hosts the domain wall. The range of stability of the transverse wall is shifted towards larger thickness compared to single Py layers, due to a magnetostatic screening effect between the two layers. An approached analytical scaling law is derived, which reproduces faithfully the phase diagram.

\end{abstract}


\maketitle

\vskip 0.5in

\vskip 0.5in


\newcommand{\Mco}{{\ifmmode{M_\mathrm{Co}}\else{$M_\mathrm{Co}$}\fi}}%
\newcommand{\Mpy}{{\ifmmode{M_\mathrm{Py}}\else{$M_\mathrm{Py}$}\fi}}%
\newcommand{\Aco}{{\ifmmode{A_\mathrm{Co}}\else{$A_\mathrm{Co}$}\fi}}%
\newcommand{\Apy}{{\ifmmode{A_\mathrm{Py}}\else{$A_\mathrm{Py}$}\fi}}%
\newcommand{\tco}{{\ifmmode{t_\mathrm{Co}}\else{$t_\mathrm{Co}$}\fi}}%
\newcommand{\tpy}{{\ifmmode{t_\mathrm{Py}}\else{$t_\mathrm{Py}$}\fi}}%
\newcommand{\tcu}{{\ifmmode{t_\mathrm{Cu}}\else{$t_\mathrm{Cu}$}\fi}}%
\newcommand{\tisoe}{{\ifmmode{t_\mathrm{isoE}}\else{$t_\mathrm{isoE}$}\fi}}%

\begin{figure}
  \begin{center}
  \includegraphics[width=87mm]{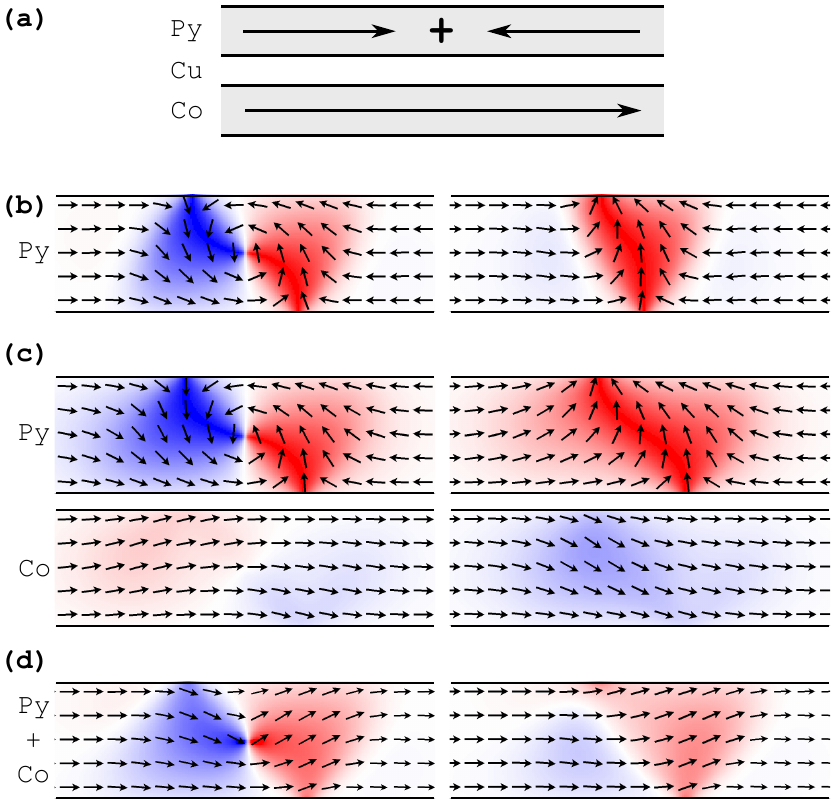}%
  \caption{\label{fig-configs}(a) Cross-sectional sketch of a head-to-head domain wall and the associated charge in a spin valve. Plane views of a vortex wall and transverse wall in a $\lengthnm{300}$-wide (b)~single layer of Py($\lengthnm{5}$) and (c)~a Co(5 nm)/Cu(5 nm)/Py(5 nm) spin valve. (d)~thickness-integrated maps of magnetization from (c), see \eqnref{eqn-integratedM}. The red (blue) color indicates the positive (negative) transverse component of magnetization, normalized to the magnetization of the layer under focus in (b) and (c), and to half the magnetization of Co in (d).}
  \end{center}
\end{figure}

The fundamental study of magnetic domain walls~(DWs) has found a playground in lithographically-defined nano-stripes. In the range of widths from a few tens to a few hundreds of nanometers their complexity is dramatically reduced with respect to extended thin films, while still retaining a few internal degrees of freedom leaving a rich physics. For stripes with in-plane magnetization, head-to-head DWs are of transverse~(TW) or vortex~(VW) type\bracketsubfigref{fig-configs}b\cite{bib-MIC1997,bib-NAK2005}, and may be characterized by a chirality, asymmetry and/or polarity. Their propagation under the stimulus of a magnetic field or a current of spin-polarized electrons is intrinsically precessional and reveals effects such as DW inertia\cite{bib-CHA2010} and a so-called Walker limit beyond which periodic DW transformations occur\cite{bib-CAY2004,bib-ALL2005,bib-BEA2005,bib-PAR2008}. Due to their small size and fast dynamics, magnetic domain walls~(DWs) are promising candidates in the area of information processing and storage\cite{bib-ALL2005,bib-PAR2008}.

Most reports so far have considered DWs in a single layer of magnetically-soft materials, such as Permalloy~(Py) or CoFeB alloys. Magnetic trilayers F1/NM/F2 (with $\mathrm{F}i$ a ferromagnet, and NM a non-magnetic material) also deserve attention as they play a key role in spintronics, especially in devices using Giant Magneto-Resistance. These stacking are called spin valve for a metal spacer layer, and pseudo spin valve or magnetic tunnel junction for an insulator spacer layer. F1 is a soft layer in which the DW moves, while F2 is a reference layer that is supposed to remain uniformly magnetized. Unusually-large current-driven DW mobilities with speed exceeding \unit[600]{m/s} have been reported in such stacks\cite{bib-GRO2003b,bib-UHL2010}. Tentative explanations point at either out-of-plane spin accumulation and torque\cite{bib-KHV2009,bib-BOO2010}, Oersted field\cite{bib-UHL2011} or magnetostatic coupling between the layers that may modify the statics or dynamics of the DW\cite{bib-NDJ2009,bib-UHL2011}.

While the role of magnetostatic coupling is well established in thin films with the formation of quasi-walls in the supposedly uniformly-magnetized layer\cite{bib-VOG2007}, it has been scarcely addressed\cite{bib-NDJ2009} in stripes. In this Letter we determine the phase diagram of head-to-head DWs in spin valves with in-plane magnetization (TW versus VW), combining numerical simulation and analytical modeling. For this purpose, we consider Co/Cu/Py nano-stripes with various widths $w$ and thickness of the individual layers $\tco$, $\tcu$ and $\tpy$. The DW is located in the Py layer, while the Co layer is initialized with uniform magnetization. We predict that the stability of TWs is enhanced towards larger Py thickness compared to single layers\cite{bib-MIC1997,bib-NAK2005}, due to the screening in the Co layer of the stray field arising from the DW in Py.

We used finite-differences micromagnetic codes, both OOMMF\cite{bib-OOMMF-report} and our home-made GL-FFT code\cite{bib-FRU2004c}. The cell size in all simulations presented here is $\unit[4\times4\times5]{nm^3}$; fo significant difference is found for a cell size $\unit[4\times4\times2.5]{nm^3}$. Effects of the finite length of the stripe in the simulations were reduced by either fixing at both ends the magnetic moments parallel to the stripe direction, or by compensating the surface magnetic charges at both ends, however with no constraint on magnetization. The length $L$ of the stripes was set equal or larger than about $20w$, for which finite-size effects are identically small under either of the above procedures. Material parameters are $\mu_{0}\Mco=\unit[1.7593]{T}$, $\Aco=\unit[30]{pJ/m}$, $\mu_{0}\Mpy=\unit[1.0053]{T}$, $\Apy=\unit[10]{pJ/m}$, and zero magnetocrystalline anisotropy in both layers. No exchange coupling between Py and Co is considered either. Different initial conditions were chosen in Py to set a TW or VW in the Py layer, while magnetization in Co is set uniform along the stripe direction; both layers are then set free to evolve during the simulation. We considered head-to-head DWs~(two domains facing each-other, see \subfigref{fig-configs}a) with no loss of generality, as tail-to-tail DWs are equivalent under time-reversal symmetry~($\vect M\rightarrow-\vect M$). Like for single layers, TWs and VWs coexist as (meta)stable states in a large range of geometrical parameters. The iso-energy curve separating the areas where either a TW or a VW is lowest in energy, was determined for each fixed $w$ through a parabolic interpolation of the energy difference for three different values of $\tpy$. We checked that this procedure yields a phase diagram of the single Py layer in very close agreement with existing reports\cite{bib-MIC1997,bib-NAK2005}.

\begin{figure}
  \begin{center}
  \includegraphics[width=85mm]{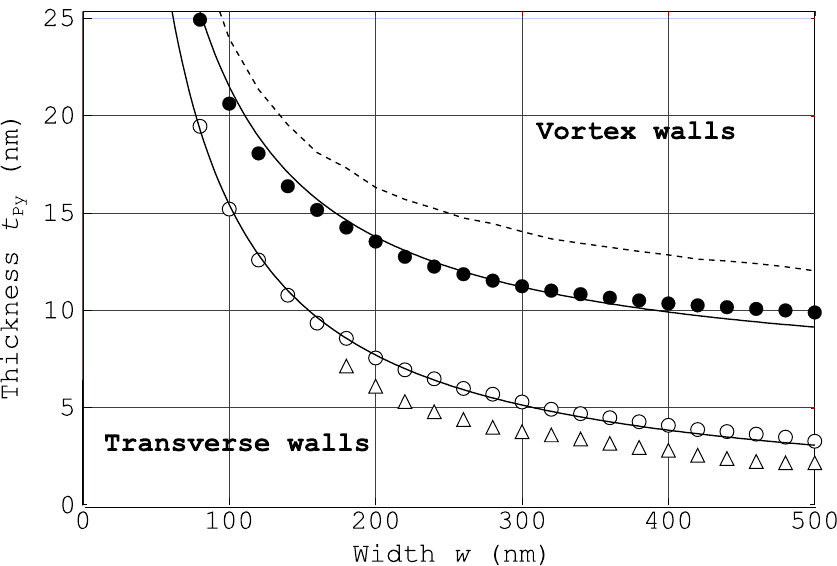}%
  \caption{\label{fig-phase-diagram}Phase diagram of head-to-head domain walls in stripes, with the boundary between symmetric and asymmetric TWs~($\triangle$), and TW and VW~($\circ$) in a single layer, and asymmetric TW and VW in a spin valve~($\bullet$). From bottom to top, the three lines are: a $1/\tpy$ fit to the TW-VW data for a single layer, a thickness translation of the former to fit the TW-VW data for a spin valve, and another thickness translation with $t_\mathrm{sh}(\tcu=0)$, see \eqnref{eqn-thickness-shift} and text for details.}
  \end{center}
\end{figure}

We first discuss qualitatively the difference between a DW in a single layer and in a spin valve. The head-to-head DW in Py is associated with a total magnetic charge $+2\tpy w\Ms$\bracketsubfigref{fig-configs}a. While VWs are similar in both systems, clear differences are found for TWs. In a single layer, TWs are symmetric for small thickness and width\bracketfigref{fig-phase-diagram}. Going towards a larger thickness or width, the TW becomes asymmetric, reducing the magnetostatic energy by spreading the magnetic charges\cite{bib-NAK2005}\bracketsubfigref{fig-configs}b; this corresponds to the onset of zig-zag walls in continuous films. The transition from symmetric to asymmetric TW is of second order~(continuous), so that for a given geometry only one type of TW exists, either symmetric or asymmetric. Asymmetric TWs are found for a large range of values of width and thickness, although they are the state of lowest energy (over the VW) only in a narrow range of values\bracketfigref{fig-phase-diagram}. In a spin valve the Co layer deviates from uniform magnetization due to the stray field arising from the DW in Py, which in turn creates a stray field acting on the Py layer. This magnetic screening effect lowers the energy compared to a single layer, as already pointed out\cite{bib-NDJ2009}. The stray field arising from the DW in Py is parallel or antiparallel to the initial direction of magnetization of Co, depending whether the left or right side of the DW\bracketsubfigref{fig-configs}a is considered. Thus, the magnetic configuration of Co is expected to be asymmetric, and so will be the stray field arising from Co and acting on the Py layer. A TW in a spin valve is therefore expected to be always asymmetric by nature, due to the unidirectional magnetization in the Co layer.

\subfigref{fig-configs}c shows a VW and a TW in a spin valve, with the same stripe width and Py thickness as in \subfigref{fig-configs}b. The magnetic screening in Co is clear from the non-uniformity of magnetization. As expected from the symmetry arguments given above the TW is asymmetric, to a larger extent than in the single layer case\cite{bib-NDJ2009}. Both the VW and TW have a larger width, associated with the emergence of tails. Indeed, as a DW profile results from the balance of magnetostatic with exchange energy, the decrease of magnetostatic energy allows to reduce exchange energy via the increase of the DW width. This is consistent with simple analytical models\cite{bib-HUB1998b} derived for charged (head-to-head) domain walls in extended spin-valve thin films.

The flux closure taking place between Py and Co can also be illustrated by the map of magnetization $M_{\mathrm{int}}$ integrated over the two layers along their normal \bracketsubfigref{fig-configs}d:

\begin{equation}
\label{eqn-integratedM}
  \vect M_{\mathrm{int}}=\frac1{\tco + \tpy}\int{\vect M(z)\diff z}
\end{equation}

\noindent This map shows similarities with a VW, which highlights the principle of reduction of the total energy: the flux is better closed than in a single layer TW, while avoiding the cost associated with the vortex core in a single layer VW. In this view the DW tails of the individual layers are also absent.

As the share of magnetostatic energy is larger in a TW than in a VW\cite{bib-MIC1997}, the inter-layer closure of flux is lowers the energy of the TW more than that of the VW. This suggests that the TW should be the ground state in a larger range of geometrical parameters in a spin valve than in a single layer. This is confirmed by a set of simulations with varying $\tpy$ and $w$, illustrated in \figref{fig-phase-diagram} for $\tco=\tcu=\lengthnm{5}$. Notice that in a single layer TWs with opposite asymmetries (left or right) are degenerate in energy, while this degeneracy is lifted in a spin valve, due to the unidirectional magnetization in the underlying Co. In the phase diagram\bracketfigref{fig-phase-diagram} we considered only the TW with the lowest energy.

We now derive a simple analytical model to grasp the main features of this diagram. McMichael and Donahue already noticed that the iso-energy line in a single layer follows the power law $tw=C\ExchangeLength^2$, with $\ExchangeLength=\sqrt{A/\Kd}$ the dipolar exchange length, $\Kd=(1/2)\muZero\Ms^2$ the dipolar constant, and $C$ a constant. This law can be derived qualitatively by balancing the energies at play in each domain wall\cite{bib-MIC1997}. These authors argued that the energy of the TW can be estimated from the lateral demagnetizing coefficient scaling like $t/w$, and the volume of the TW of the order of $tw^2$ (thus $t^2 w$ as a whole), while that of the VW is related to the energy of the vortex core, scaling with $t\ExchangeLength^2\Kd$. The exchange energy in the $\approx\angledeg{90}$ N\'{e}el sub-walls can be ignored as the total length of these walls is identical in a TW and a VW. The numerical value $C\approx61-64$ must be provided by simulations\cite{bib-MIC1997,bib-NAK2005}.

This scaling model can be adapted to a spin valve, however the above arguments must first be discussed in more detail. It shall first be noted that in any head-to-head DW, either TW or VW, most of the energy is of magnetostatic origin and related to the head-to-head charges, so that the above argument must be refined. Simulation and magnetic force microscopy of a head-to-head DW in a single layer show that the total charge $Q=2tw\Ms$ is nearly uniformly spread over the area of the DW\cite{bib-CHA2010}, which is $\approx w^2$ for a TW, and $\approx 2w^2$ for a VW. The associated densities of surface charges are then $\sigma_\mathrm{TW}=2(t/w)\Ms$ and $\sigma_\mathrm{VW}=(t/w)\Ms$. The resulting volume density of stray field energy scales with $\sigma^2$ and extends over a typical distance $w$ above and below the stripe. This results in an excess of magnetostatic energy in the TW over the VW of $\approx2t^2w\Kd$. We now consider a spin valve. In the limit $\tpy\Mpy>\tco\Mco$, which is suitable for $\tco=\lengthnm{5}$ and the boundaries of the phase diagram of practical interest\cite{bib-MIC1997,bib-NAK2005}, the deformation in Co reaches its maximum while not being enough for a full screening of the charges arising from the DW in Py. The above scaling laws can be rewritten accordingly and the following law is derived again with some approximations (see below), making also use of the partial flux closure in spin valves to refine the lateral demagnetizing coefficient\cite{bib-FRU2012}:

\begin{equation}
\label{eqn-scalingLaw}
w(t-t_\mathrm{sh})\approx C\ExchangeLength^2.
\end{equation}

\noindent This suggests that the phase diagram for a spin valve is based on that of a single layer, shifted towards higher thicknesses by a value $t_\mathrm{sh}$. This is in very good agreement with the results of numerical simulations\bracketfigref{fig-phase-diagram}. Quantitatively, the analytical formula for $t_\mathrm{sh}$ is not simple, implying both power and logarithmic functions. However, thanks to the slow variation of the latter, $t_\mathrm{sh}$ can be approximated faithfully with $\tco(\Mco/\Mpy)\approx1.75\tco$.

\begin{figure}
  \begin{center}
  \includegraphics[width=73.207mm]{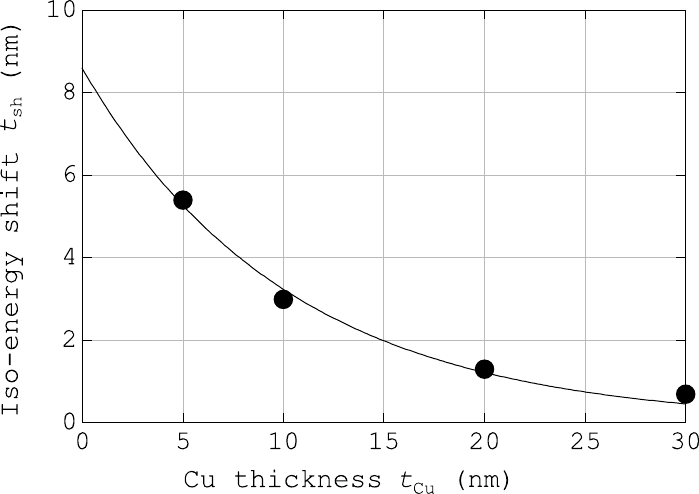}%
  \caption{\label{fig-cuspacer}Thickness shift of the iso-energy line in a spin valve compared to a single layer, for $w=\lengthnm{100}$ and $\tco=\lengthnm{5}$.}
  \end{center}
\end{figure}

For $\tco=\tcu=\lengthnm{5}$ the analytical model thus yields $1.75\tco\approx\lengthnm{8.75}$, while $t_\mathrm{sh}$ determined from simulations is $\thicknm{5.4}$. This discrepancy is linked with the magnetic screening in the Co layer being partial. This can be understood as in the scaling model we neglect the magnetostatic energy stored in the volume of the Cu spacer, locus of a magnetic field arising between the charges of opposite sign in Py and in Co. \figref{fig-cuspacer} shows $t_\mathrm{sh}$ resulting from micromagnetic simulations as a function of $\tcu$ for $w=\lengthnm{100}$ and $\tco=\lengthnm{5}$. For large $\tcu$ the problem tends towards the case of a single layer because the screening is no more effective. The decay is close to exponential, with a fit providing the extrapolation $t_\mathrm{sh}(\tcu=0)\approx\lengthnm{8.60}$. This figure is now in excellent agreement with the analytical model, which suggests the following empirical law:

\begin{equation}
\label{eqn-thickness-shift}
t_\mathrm{sh}=\tco\frac{\Mco}{\Mpy}\mathrm{e}^{-\frac{\tcu}{t_0}}
\end{equation}

\noindent where $t_0\approx\lengthnm{10}$ is derived from \figref{fig-cuspacer}. Additional micromagnetic simulations show that $t_0$ increases with $w$, \eg $t_0=\lengthnm{13}$ and $\lengthnm{16}$ for $w=\lengthnm{300}$ and $\lengthnm{500}$, respectively. This trend can be understood as the initial stray field (to be reduced through screening) arising from the DW extends outside the stripe over a volume scaling with $w^3$, while the magnetic field arising between the charges of opposite sign in Py and in Co (a side cost of the screening effect) applies in a volume scaling with $\tcu w^2$. Thus, the relative gain of energy through screening decreases with decreasing the stripe width. Then, \eqnref{eqn-thickness-shift} can be applied for other Co thicknesses, or more generally to other couples of magnetic materials if substituting the proper spontaneous magnetization values in \eqnref{eqn-scalingLaw}, and fitting the decay of $t_\mathrm{sh}$ like in \figref{fig-cuspacer}.

One must finally keep in mind that for $\tpy\Mpy<\tco\Mco$ a very effective screening of the head-to-head DW charge is possible, so that deviations from \eqnref{eqn-scalingLaw} are expected. In that case, the energy of the TW is drastically reduced. The iso-energy line becomes very flat beyond this limit, an effect that starts to be visible for the largest widths in \figref{fig-phase-diagram}.

To conclude, we have investigated with numerical simulation and analytical scaling laws the phase diagram of head-to-head domain walls in F1/NM/F2 spin valves, with a domain wall in one layer and no domain wall in the other layer. We showed that the range of stability of transverse versus vortex walls is enhanced due to a magnetic screening effect. The iso-energy line in spin valves is translated towards a larger thickness with respect to single layers, by a value decreasing approximately exponentially with the spacer thickness. This enhanced stability and larger width provide a magnetostatic contribution to the high mobility of DWs in trilayers, provided that the damping in Co is not too large\cite{bib-NDJ2009}. Effects on the domain wall inertia and automotion\cite{bib-CHA2010} are also expected.

\section*{References}

\section*{Acknowledgements}

This work was partially supported by ANR-07-NANO-034 \textsl{Dynawall} and ANR-08-BLAN-0199 \textsl{MicroManip}. VU was supported by the research programmes of GAAV (Project No. KAN400100701) and European Regional Development Fund (CEITEC - CZ.1.05/1.1.00/02.0068).

\end{document}